\documentclass[twocolumn]{aastex631}

\begin{document}

\title{Early JWST imaging reveals strong optical and NIR color gradients in galaxies at $z\sim2$ driven mostly by dust}

\correspondingauthor{Tim B. Miller}
\email{tim.miller@yale.edu}

\author[0000-0001-8367-6265]{Tim B. Miller}
\affiliation{Department of Astronomy, Yale University, 52 Hillhouse Ave., New Haven, CT, USA, 06511}

\author[0000-0001-7160-3632]{Katherine E. Whitaker}
\affil{Department of Astronomy, University of Massachusetts, Amherst, MA 01003, USA}
\affiliation{Cosmic Dawn Center (DAWN), Niels Bohr Institute, University of Copenhagen, Jagtvej 128, K\o benhavn N, DK-2200, Denmark}

\author[0000-0002-7524-374X]{Erica J. Nelson}
\affiliation{Department for Astrophysical and Planetary Science, University of Colorado, Boulder, CO 80309, USA}

\author[0000-0002-8282-9888]{Pieter van Dokkum}
\affil{Department of Astronomy, Yale University, 52 Hillhouse Ave., New Haven, CT, USA, 06511}

\author[0000-0001-5063-8254]{Rachel Bezanson}
\affiliation{Department of Physics and Astronomy and PITT PACC, University of Pittsburgh, Pittsburgh, PA 15260, USA}

\author[0000-0003-2680-005X]{Gabriel Brammer}
\affiliation{Cosmic Dawn Center (DAWN), Niels Bohr Institute, University of Copenhagen, Jagtvej 128, K\o benhavn N, DK-2200, Denmark}

\author[0000-0002-9389-7413]{Kasper E. Heintz}
\affiliation{Cosmic Dawn Center (DAWN), Niels Bohr Institute, University of Copenhagen, Jagtvej 128, K\o benhavn N, DK-2200, Denmark}

\author[0000-0001-6755-1315]{Joel Leja}
\affiliation{Department of Astronomy \& Astrophysics, The Pennsylvania State University, University Park, PA 16802, USA}
\affiliation{Institute for Computation \& Data Sciences, The Pennsylvania State University, University Park, PA 16802, USA}
\affiliation{Institute for Gravitation and the Cosmos, The Pennsylvania State University, University Park, PA 16802, USA}

\author[0000-0002-1714-1905]{Katherine A. Suess}
\affiliation{Department of Astronomy and Astrophysics, University of California, Santa Cruz, 1156 High Street, Santa Cruz, CA 95064 USA}
\affiliation{Kavli Institute for Particle Astrophysics and Cosmology and Department of Physics, Stanford University, Stanford, CA 94305, USA}

\author[0000-0003-1614-196X]{John R. Weaver}
\affil{Department of Astronomy, University of Massachusetts, Amherst, MA 01003, USA}

\begin{abstract}
Recent studies have shown that galaxies at cosmic noon are redder in the center and bluer in the outskirts, mirroring results in the local universe. These color gradients could be caused by either gradients in the stellar age or dust opacity; however, distinguishing between these two causes is impossible with rest-frame optical photometry alone. Here we investigate the underlying causes of the gradients from spatially-resolved rest-frame $U-V$ vs.\ $V-J$ color-color diagrams, measured from early observations with the \emph{James Webb Space Telescope}.  We use $1\, \mu m - 4\, \mu m$ NIRCam photometry from the CEERS survey of a sample of 54 galaxies with $\log M_* / M_\odot>10$ at redshifts $1.7<z<2.3$ selected from the 3D-HST catalog. We model the light profiles in the F115W, F200W and F356W NIRCam bands using \texttt{imcascade}, a Bayesian implementation of the Multi-Gaussian expansion (MGE) technique which flexibly represents galaxy profiles using a series of Gaussians. We construct resolved rest-frame $U-V$ and $V-J$ color profiles. The majority of star-forming galaxies have negative gradients (i.e. redder in the center, bluer in the outskirts) in both $U-V$ and $V-J$ colors consistent with radially decreasing dust attenuation. A smaller population (roughly 15\%) of star-forming galaxies have positive $U-V$ but negative $V-J$ gradients implying centrally concentrated star-formation. For quiescent galaxies we find a diversity of UVJ color profiles, with roughly one-third showing star-formation in their center. This study showcases the potential of \emph{JWST} to study the resolved stellar populations of galaxies at cosmic noon. 
\end{abstract}


\section{Introduction} \label{sec:intro}

Measuring and modelling the colors of stellar populations provides great insight into their physical nature. In the local universe, galaxies are known to have gradients in their radial color profile, indicating complex multi-component formation histories. Observed in the local universe since the late 1980s \citep{Kormendy1989, Franx1990}, both star-forming and quiescent galaxies (or late and early types) display negative color gradients (i.e. redder in the center, bluer in the outskirts). In star-forming galaxies, like the Milky Way, this is usually due to multiple stellar populations, i.e. an old red bulge and a young blue disk. In quiescent, or early type galaxies, the cause is thought to be lower metallicity in the outskirts \citep{Wu2005,Tortora2011} with the additional possibility of an age gradient \citep{LaBarbera2009}.

Recent studies have found similar results at $z>1$, with negative optical color gradients in all type of galaxies. These gradients have either been observed directly \citep{Wuyts2012,Szomoru2013,Liu2016,Wang2017, Liu2017, Miller2022} or implicitly through comparing mass-weighted radii to light-weighted radii \citep{Chan2016,Suess2019a,Mosleh2020}. Mass-to-light ratio is correlated with optical color so the fact that mass-weighted radii are smaller than light-weighted radii also implies negative color gradients. At $z>1$ spaced-based imaging is needed to resolve the structure of galaxies. Until very recently these studies relied on \emph{Hubble Space Telescope (HST)} imaging, but it is limited by its longest observable wavelength of $1.6\ \mu m$, which traces the rest-frame optical at $z>1$. 

A limitation imposed by the lack of high resolution IR imaging is the inability to understand the causes of these color gradients. Star formation history, dust attenuation and, to a lesser extent, metallicity and nebular emission lines all contribute to the optical colors of galaxies producing degeneracies between these parameters when interpreting optical colors alone. Longer wavelength measurements, especially in the near-infrared, are often used to break these degenericies. In particular the plane of rest frame $U-V$ and $V-J$ colors, the UVJ diagram, is commonly used.~\citep{Labbe2005,Wuyts2007} In this plane the effects of dust and star-formation history on optical colors can be separated. Galaxies with high Dust attenuation follows the attenuation curve and are red in both $U-V$ and $V-J$ colors where as old stellar populations are relatively bluer in $V-J$ colors.

The UVJ diagram \citep[or other similar rest-frame color-color selections, e.g.][]{Arnouts2013,Leja2019,AntwiDanso2022} have become ubiquitous to distinguish between star-forming and quiescent galaxies \citep{Williams2009,Brammer2009,Whitaker2011,Muzzin2013,Ilbert2013,Davidzon2017,Akins2022}. While rest-frame color selections have been used when studying integrated galaxy properties, the comparably worse sensitivity and spatial resolution of \emph{Spitzer/IRAC} (i.e., the only instrument able to image rest-frame J band at $z\gtrsim 1$) has limited resolved studies of high redshift galaxies.

The landscape of spatially resolved studies of high redshift galaxies has changed with the recent launch and commissioning of the \emph{James Webb Space Telescope (JWST)}. With the combination of the longer wavelength coverage and improved spatial resolution and sensitivity, JWST facilitates studying the spatially resolved rest-frame NIR emission of galaxies at $z\sim 2$. In this paper, we utilize early release observations from the Cosmic Evolution Early Release Science (CEERS) \citep[PI: Finkelstein][]{Finkelstein2017} to investigate the question: What is the physical cause of optical color gradients at $z\sim 2$? We construct resolved UVJ diagrams of galaxies from the 3D-HST catalog and study their radial gradients of galaxies within this plane. To model the light profiles we use \texttt{imcascade}, a Bayesian implementation of the MGE technique which provides a more flexible representation compared to traditional Sersic profile fitting~\citep{Miller2021}.

This paper is organized as follows: In Section~\ref{sec:methods} we describe the reduction of the JWST images, the \texttt{imcascade} modeling procedure, along with the conversion from observed filters to rest-frame UVJ measurements. Section~\ref{sec:results} displays our resolved UVJ measurements. The physical implications of our findings are discussed in Section~\ref{sec:disc}. Throughout this study we assume a flat $\Lambda\rm CDM$ cosmology with $H_0 = 70\ \rm km / s/ Mpc$ and $\Omega_m = 0.3$. All radii are referred to along the semi-major axis. All magnitudes are reported on the AB system

\section{Methods} \label{sec:methods}

\subsection{Data, Galaxy Sample and \texttt{imcascade} Modeling}
The NIRCam imaging used in this study  were taken between June 22-28, 2022 as part of the CEERS survey ~\citep[see][]{Finkelstein2017,Finkelstein2022} covering roughly 40 sq. arcminutes in the AEGIS field. For this study we focus on the three broadband filters: F115W, F200W and F356W. Stage 2 calibrated data were used from the available MAST archive. Additional reduction, aligning and co-adding were performed with the \texttt{grizli} software package \citep{grizli}. Zeropoints were used from the available calibration file \texttt{jwst\_ 0942.pmap} with additional per-chip corrections in each band based on standard stars in the LMC\footnote{For more details on these corrections see: https://github.com/gbrammer/grizli/pull/107}. This correction has been further verified to be accurate to $<0.04$ mag using data from M92, although there may be additional time variability~\citep{Boyer2022}. All fitting was done on a common pixel scale of 0.04 $''/\rm pix$. Weight maps were also calculated using \texttt{grizli} with Poisson noise included.

To construct our galaxy sample we begin with the 3D-HST catalog in the AEGIS field ~\citep{Skelton2014,Momcheva2016} and select galaxies with $\log\ M_*/M_\odot > 10$ at $1.7<z <2.3$ that lie within the footprint of the first epoch of CEERS data.\added{ $z_{\rm best}$ is used as the redshift measurement which corresponds to the spectroscopic, grism, or photometric redshift in this rank order, depending on what is available. The grism and the photometric redshifts are calculated using \texttt{EAZY}\citep{Brammer2008} } This creates an initial sample of 119 galaxies.

We fit the light distribution for galaxies in this sample in each of the three NIRCam bands using \texttt{imcascade} \citep{Miller2021}. \texttt{imcascade} is a Bayesian implementation of the Multi-Gaussian Expansion (MGE) technique which models a galaxy light distribution as a mixture of Gaussians. This provides a more flexible representation compared to traditional parametric S\'{e}rsic fits, making it ideal for measuring complex color profiles. We briefly describe our procedure below but refer the reader to \citet{Miller2021} for a full description of the method and implementation.

To begin, we create cutouts of each galaxy in each band with a size $35\times r_{\rm vdW12}$ based on the F160W measured size from the \citet{vanderWel2012} catalog. PSFs for each band are generated using \texttt{webbpsf} \citep{webbpsf} and the drizzled PSF was calculated using the same parameters used to create the mosaic. For use in \texttt{imcascade}, we fit the PSF in each band with an Multi-Gaussian expansion (MGE) model using 5 Gaussian components for F115W and F200W and 4 components for F356W. We use the same set of 10 Gaussian components to model each galaxy in every band, with widths logarithmically spaced from 0.75 pixels to  $9\times r_{\rm vdW12}$. In each image a 3 parameter tilted plane sky model is simultaneously fit. Masks for nearby sources are created for each band separately using the segmentation tool in \texttt{photutils}~\citep{photutils}, specifically using a signal-to-noise threshold of 3 with a de-blending threshold of 0.005. We expand this initial mask (where masked pixels have a value of 1) by convolving with a Gaussian of width 2.5 pixels and masking all pixels above 0.01.

Bayesian inference with \texttt{imcascade} is carried out with the nested sampling code \texttt{dynesty} \citep{Speagle2020} utilizing the `express' method where the position, axis ratio and position angle of the components are measured from least-squares fitting and are kept constant while the posteriors of the fluxes of each component and the sky parameters are explored. This allows for much faster model creation greatly speeding up execution time by a factor of roughly 100. The fluxes of each component are explored in logarithmic space using the informed priors discussed in \citet{Miller2021}.\added{Throughout the paper we report colors integrated between in annular bins, using the best fit axis ratio and position angle, from the intrinsic (i.e. PSF deconvolved) \texttt{imcascade} models for each filter of each galaxy. For each measurement we take 150 samples from the posterior distribution and report the median value with error bars representing the 16th-84th percentile range. }

We perform a set of quality checks once inference is complete to select a high-quality sample for this study. In each filter we ensure that none of the parameters in the optimized least-squares solution are at the bounds provided, a sign of an ill-converged fit and that the derived parameters are not robust. Additionally we require the axis ratio to $q>0.1$ and the flux contained in the \replaced{largest component}{Gaussian component with the largest width} to be less than 20\%  of the total flux, both signs of issues with background structure or unmasked sources. We find 54 of the 119 galaxies pass all of the checks and will be used for the remainder of the paper. The distribution of redshift, stellar mass and specific star-formation rate for the galaxies which did and did not pass these checks are similar. From visual inspection, we find that most of the galaxies which did not pass had issues with masking. Some were present in crowded fields where proper masking is difficult, while others had the target galaxy over-masked or nearby sources left unmasked.

\subsection{Conversion to Rest-Frame Filters}
\begin{figure*}
    \centering
    \includegraphics[width = 0.9\textwidth]{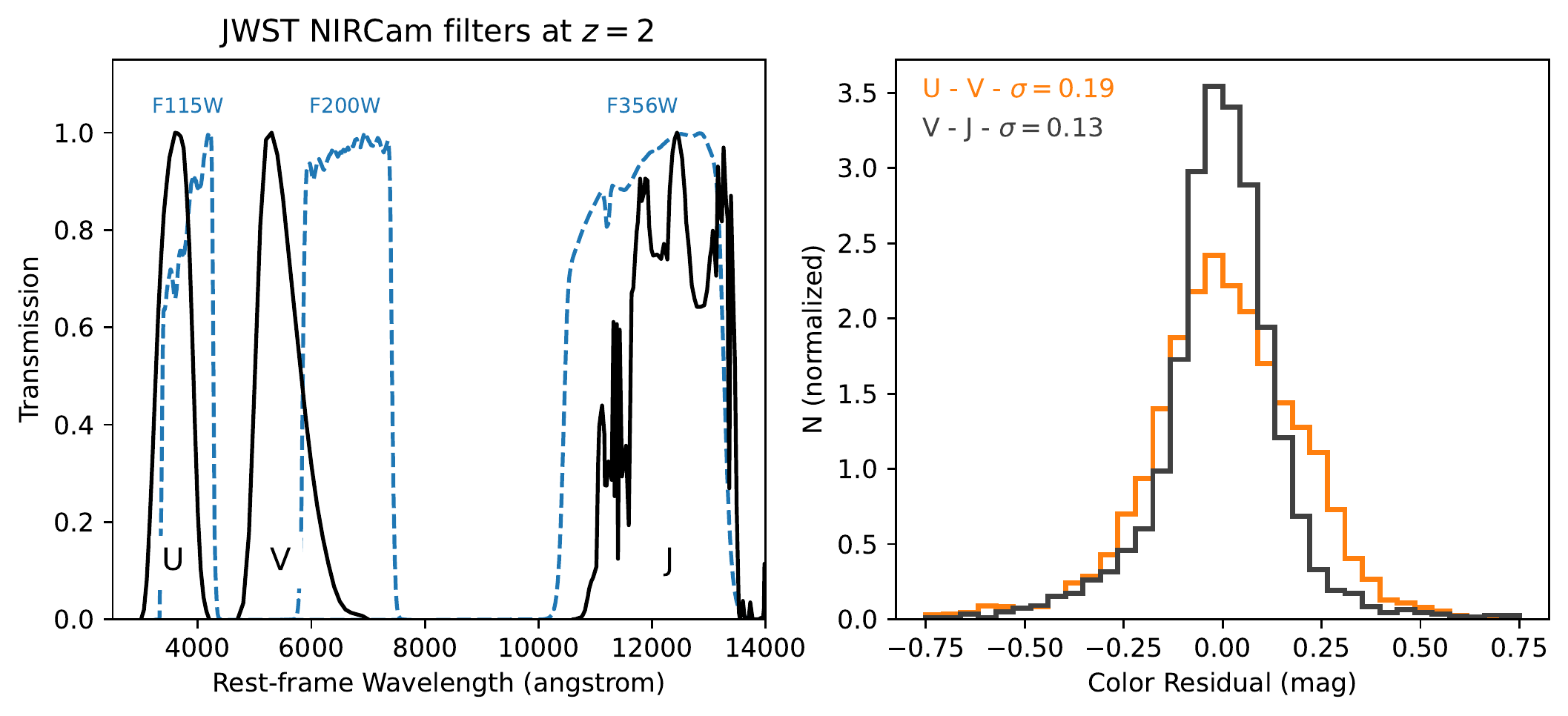}
    \caption{\textit{(left)} Filter transmission curves for rest-frame U,V and J (black) along with JWST filters for a galaxy if observed at z=2 (Blue dashed).  \textit{(right)} Residuals between true U-V and V-J colors and those calculated from observed JWST filters using Eqn 1, for $10^4$ simulated spectra with \texttt{prospector}. We find good agreement between the true and calculated rest-frame colors over the entire redshift range for all galaxy types.}
    \label{fig:filt_conv}
\end{figure*}

In order to ensure an accurate conversion between the observed and rest-frame colors we use simulated spectra to fit a redshift dependent relation. Figure~\ref{fig:filt_conv} shows normalized filter transmission curves of rest-frame $U,V,J$ compared to JWST NIRCam filters for a galaxy observed at $z=2$. For this study we use F115W as a proxy for the rest-frame $U$-band, F200W as a proxy for rest-frame $V$-band and F356W for $J$. To ensure accurate conversion we simulate $10^4$ spectra for realistic galaxies at $z\sim 2$ using \texttt{PROSPECTOR}~\citep{Leja2017,Johnson2021}. These galaxies are generated at $1.7<z<2.3$ using delayed-$\tau$ parametric star-formation histories with an additional burst component and contribution from nebular emission lines. The prior on the strength of the burst is uniform 0 to 50\% of the total stars formed and the age as a fraction of the age of the universe at $z\sim2$, is uniform between 0.5 and 1. \added{A fixed \citet{Calzetti2000} dust curve is used and the optical depth in the $V$ band is varied uniformly between 0 and 3.} From these simulated spectra we can calculate the observed flux in the JWST NIRCam filters and compare directly with the rest-frame $U,V$ and $J$-band fluxes. We fit a linear relation between observed colors and the known rest-frame colors along with a linear redshift evolution term. The form of this equation, along with the best fit values for each color are shown below:
\begin{equation}
    \small
    (U-V)_{RF} = 0.971\, (m_{115W} - m_{200W}) + 0.056 - 0.969\, (z-2)
    \label{eqn:UmV_conv}
\end{equation}
\begin{equation}
    \small
    (V-J)_{RF} = 1.310\, (m_{200W} - m_{356W}) + 0.168 - 0.268\, (z-2) 
    \label{eqn:VmJ_conv}
\end{equation}
These relations are valid only across the redshift range of $1.7<z<2.3$.  At $z<1.7$ F115W shifts redwards of the rest-frame $U$-band introducing systematic errors due to extrapolation. At $z>2.3$, F356W becomes too ``blue'' and no longer overlaps with rest-frame $J$, similarly complicating the conversion. The residuals between the true and calculated UVJ colors for the simulated galaxies are shown in Figure~\ref{fig:filt_conv}. The standard deviation of the residuals, $\sigma = 0.19\ \rm mag$ for $U-V$ and $\sigma = 0.13$ for $V-J$, is consistent across all galaxy types and is relatively small compared to the 2 mag range that galaxies span in each color. This scatter is consistent with redshift for the $V-J$ conversion but increases slightly at lower redshifts for $U-V$. While typical \deleted{photometric} redshift uncertainties ($\sigma_z \lesssim 0.05$) will not increase this scatter significantly, catastrophic errors in redshifts of individual galaxies will lead to outliers in rest-frame color space. This is only expected for roughly 3\% of galaxies in the AEGIS field~\citep{Skelton2014,Bezanson2016}.

\added{We follow previous studies and use a simple relation to convert to rest-frame colors because we are only using 3 filters~\citep{Wang2017}. However, there are methods, such as EAZY~\citet{Brammer2008}, which use physically motivated templates for the galaxy's SED which should lead to a more accurate conversion. To test if there is any benefit, we apply EAZY to calculate the rest-frame colors from the \texttt{PROSPECTOR} mock photometry using the same 3 filters. We find very similar accuracy to the linear relations used here, the standard deviation of EAZY residuals are 0.21 mag for $U-V$ and 0.12 for $V-J$, compared to 0.19 and 0.13 respectively for the linear relations. While this is not a perfect comparison as EAZY and \texttt{PROSPECTOR} make different assumptions about the physical properties of galaxies, it is indicative that the template based methods provide little benefit over the simple linear relation when a small number of filters are used.}

As a check on our procedure we compare our rest-frame color measurements to those in the 3D-HST catalog. We integrate our \texttt{imcascade} models convolved with the PSF to a radius of 0.35 arcsec, similar to the apertures used in the 3D-HST catalog~\citep{Skelton2014}. We find good agreement, with a mean difference of less than 0.05 mag and a scatter of $0.23$ mag for both colors. 

\section{Results} \label{sec:results}
\begin{figure*}
    \centering
    \includegraphics[width = 0.99\textwidth]{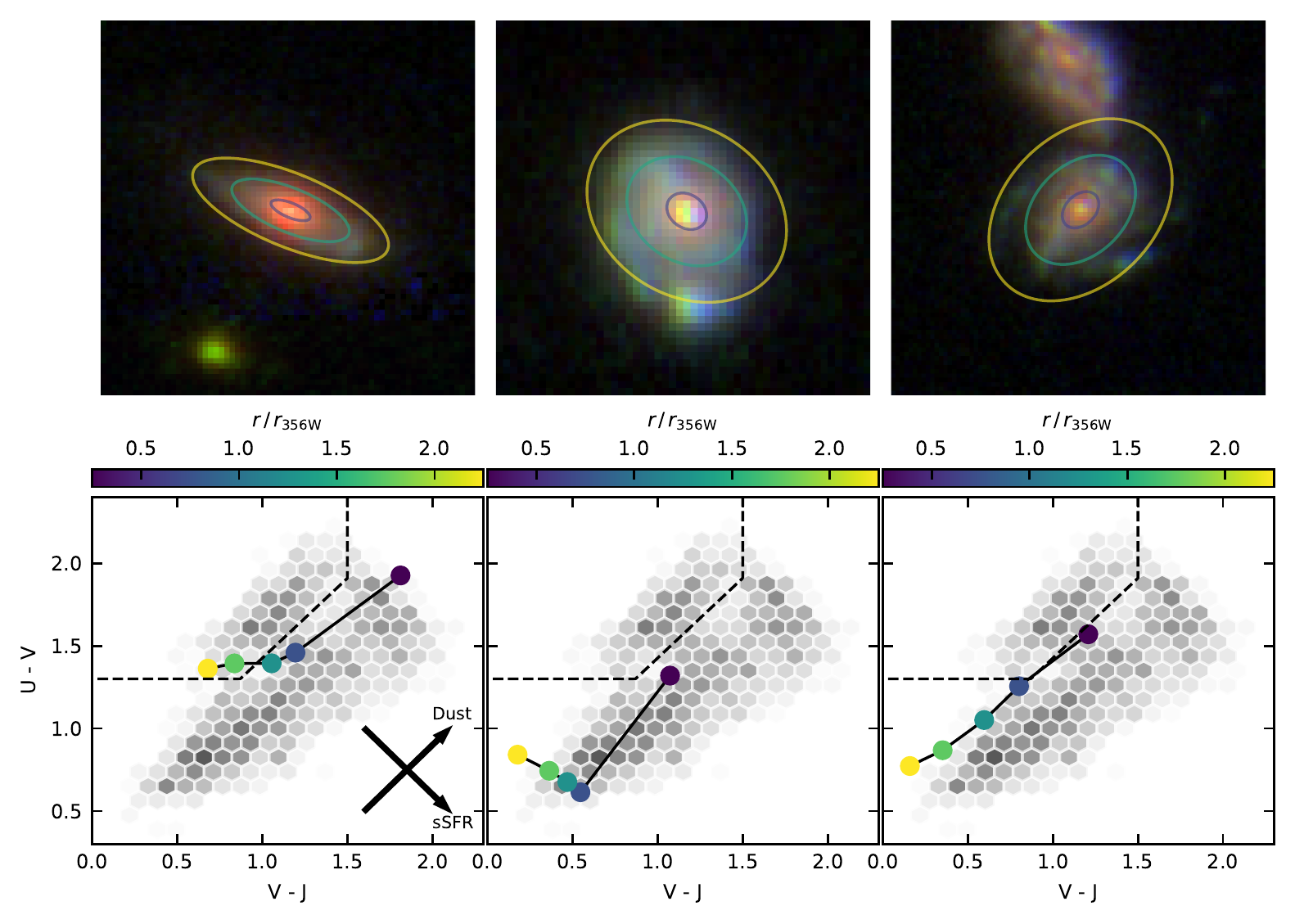}
    \caption{RGB images (F356W, F200W and F150W, filters) of three example galaxies (top) and their corresponding resolved UVJ diagrams (bottom).  The colored track indicates the evolution in rest-frame color from the innermost region (purple) to the outskirts at $>2r_{e}$ (yellow). These colors are mirrored on the RGB images to show the corresponding radii. The dotted line shows the cutoff between star-forming and quiescent from \citet{Muzzin2013} and the grey histogram in the background shows the distribution of galaxies in the same mass and redshift range from 3D-HST. Vectors illustrating the expected effect of increasing dust and sSFR are indicated in the bottom right corner.}
    \label{fig:examp_gals}
\end{figure*}

RGB renderings showing F356W, F200W and F150W images of three example galaxies are shown in Figure~\ref{fig:examp_gals}. We construct resolved UVJ diagrams by calculating the integrated colors with bins of width $0.5\, r_{F356W}$ from the center out to $2.5\, r_{F356W}$. The colors of each point denote the average radii used to calculate the UVJ colors and correspond to the ellipses plot on the RGB images. These measurements are shown alongside the distribution of integrated UVJ colors from galaxies over the same mass and redshift range from the 3D-HST survey. The three galaxies chosen are all at $z\sim1.8$ and show a variety of UVJ color profiles. 

To help build intuition we show illustrative arrows indicating how we expect specific star-formation rate (sSFR) and dust to vary across the UVJ plane. The effect of dust follows a vector in this plane set by the attenuation curve. A \citet{Calzetti2000}-like dust curve is often used but there is known to be some variation in the attenuation curves of galaxies at high redshift~\citep{Kriek2013,Reddy2015}. Following this relation, dust attenuation increases from bottom-left to top-right. The sSFR is observed to vary in the orthogonal direction, decreasing from bottom right to top left \citep{Wang2017}. Quiescent galaxies are commonly chosen as being above the selection line shown in Fig~\ref{fig:examp_gals} \citep{Muzzin2013}. In essence, this is selecting for the presence of a strong Balmer break, indicative of an old stellar population. Within this quiescent region there is also known to be an age sequence \citep{Whitaker2012,Whitaker2013,Belli2019}. We discuss the interpretation of quiescent galaxies further in Section~\ref{sec:res_Q}. We refer the reader to \citet{Leja2019} for a more in depth discussion of how galaxy properties vary across the UVJ plane.

The three galaxies shown all display disky morphology with a red center. The galaxies on the left and right are bluer in $U-V$ and $V-J$ at larger radii consistent with radially decreasing dust attenuation \citep{Nelson2016,Liu2017}. The galaxy on the left appears to be edge-on and has much higher dust attenuation overall~\citep[e.g][]{Nelson2022}. The galaxy in the center also shows high central dust attenuation but shows positive $U-V$ gradient at $r>r_{\rm 356W}$. This implies higher sSFR in the outskirts. These images display the ability of \emph{JWST} to study the resolved structure of galaxies at cosmic noon along with the complex galaxy structure already in place at the epoch. Comparison of the UVJ gradients highlight the need for sub-arcsecond rest-frame NIR observations to untangle the spatially complex dust and stellar populations of $z\sim2$ galaxies..

\begin{figure*}
    \centering
    \includegraphics[width = 0.95\textwidth]{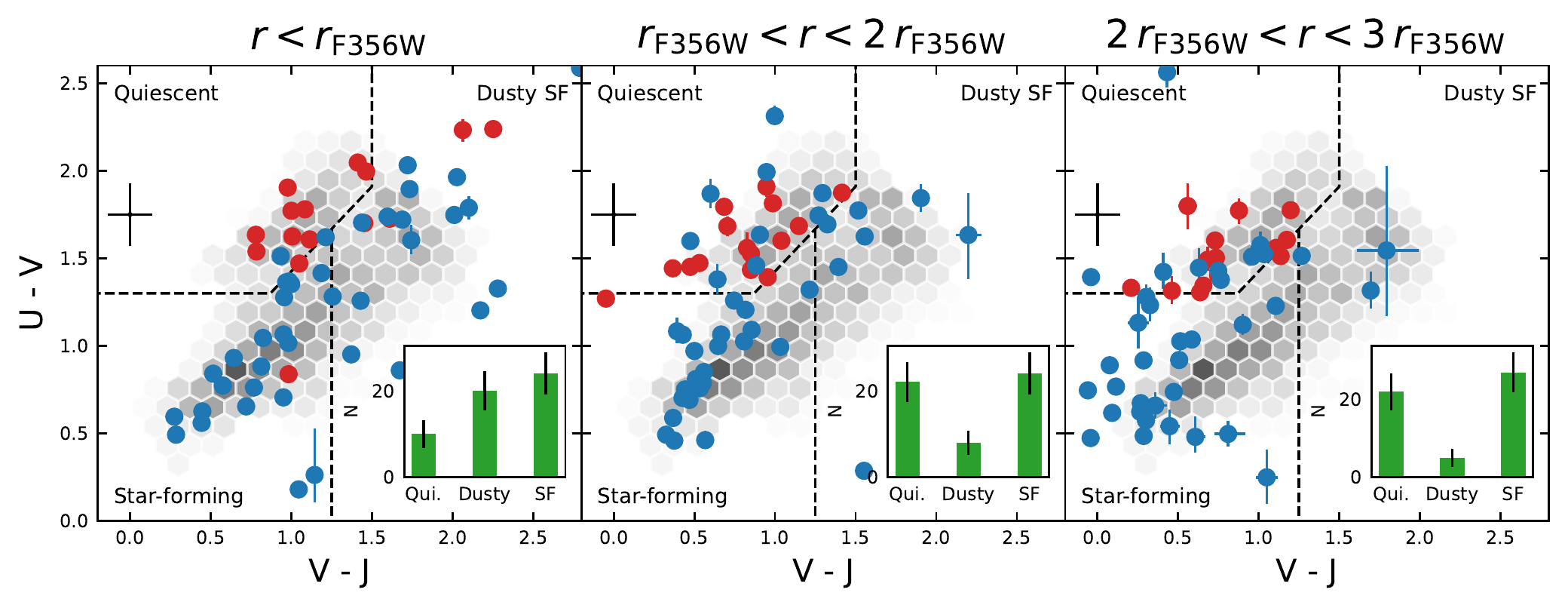}
    \caption{UVJ colors for individual galaxies measured within three separate radial bins. The dotted line shows the cutoff between star-forming and quiescent from \citet{Muzzin2013} and the grey histogram in the background shows the distribution of galaxies in the same mass and redshift range from 3D-HST. Galaxies are separated into star-forming (blue) and quiescent (red) based on their total U-V and V-J colors measured with \texttt{imcascade}. Observational uncertainties are shown as errorbars on individual points and the black errorbar in the top left shows the additional uncertainty due to converting from observed colors to rest-frame colors. We classify each region of each galaxy into quiescent, star-forming or dusty or star-forming (see text for more details) and the inset bar charts show the distribution of each classification at each radii.}
    \label{fig:resolved_uvj}
\end{figure*}

Figure~\ref{fig:resolved_uvj} displays a resolved view of the UVJ color-color plane for the galaxies in our sample. The three panels show UVJ colors measured from the intrinsic \texttt{imcascade} models integrated between $0<r<r_{\rm F356W}$, $r_{\rm F356W)}<r<2\,r_{\rm F356W}$ and  $2r_{\rm F356W}<r<3\,r_{\rm F356W}$, respectively from left to right.  At small radii we find some star-forming galaxies with very strong dust attenuation ($U-V \gtrsim 1.5, V-J \gtrsim  1.5$) but this population largely disappears when looking at the outskirts of galaxies. The entire population of star-forming galaxies appears to shift towards the bottom left of the color-color plane i.e. predominantly unobscured star formation. The behaviour and interpretation of quiescent galaxies in this plane is more complex and is discussed further in Section~\ref{sec:res_Q}.

\added{To help understand the physical causes of these color gradients we classify each radial bin of each galaxy as quiescent, dusty star-forming or star-forming}. The quiescent selection follows \citet{Muzzin2013}. We then classify each radial bin of each galaxy as dusty-star forming if $V-J >1.25$ and as star-forming if not. We note that this classification may not be intrinsic and is impacted by viewing angle and dust geometry.~\citep{Patel2012,Zuckerman2021} These regions are labelled on Fig~\ref{fig:resolved_uvj} and \added{the inset bar graph displays the distribution of classifications at each radii.}

We find the number of galaxies classified as star-forming remains constant in all three radii bins. The number galaxies classified as dusty star-forming is high in the inner regions but drops significantly in the outer regions. Correspondingly, when looking at the inner regions there are only 10 galaxies classified as quiescent in their inner region, which rises to 22 at $2r_{\rm F356W}<r<3\,r_{\rm F356W}$. In this largest radii bin there are a number of galaxies that have UVJ colours near the cutoff between star-forming and quiescent with $U-V \lesssim 1.5$ and $V-J< 1$.

\begin{figure*}
    \centering
    \includegraphics[width = 0.99\textwidth]{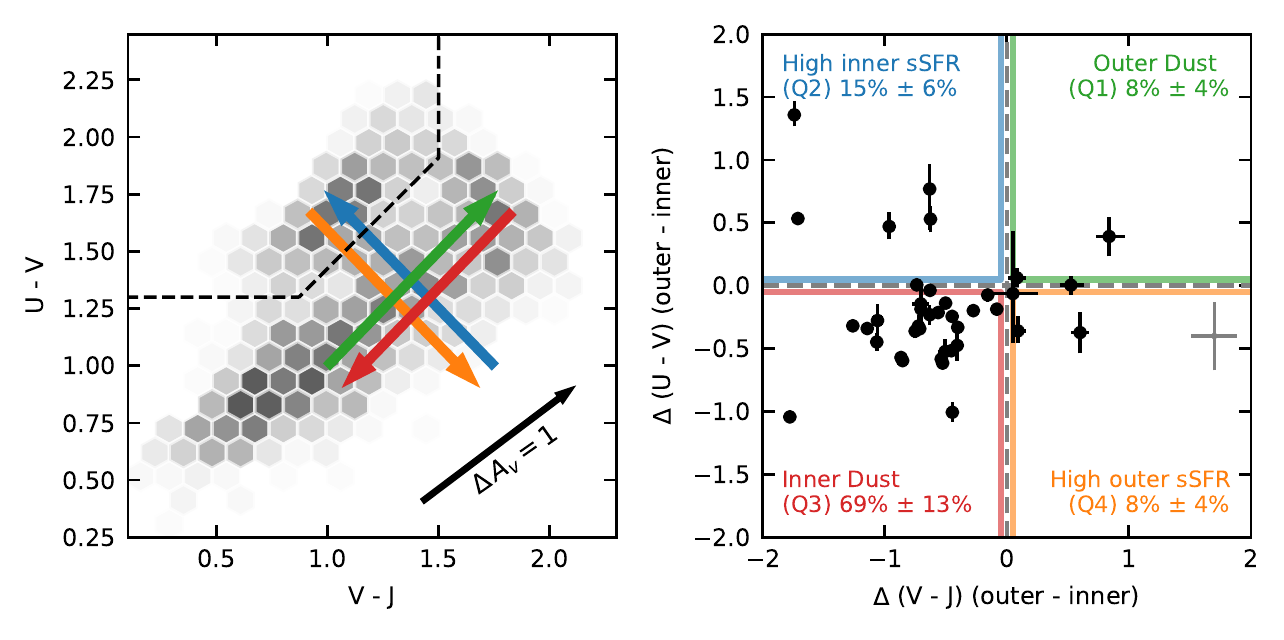}
    \caption{UVJ gradients of star-forming galaxies. \textit{(right)} The change in the U-V and V-J colors from the outer galaxy ($2\, r_{\rm 356W} < r < 3\, r_{\rm F356W}$) to the inner galaxy ($ r < \, r_{\rm F356W}$) is shown. As with the figure above observational uncertainties are shown on individual data points and the additional uncertainty due to converting from observed to rest frame filts is shown as in the grey errorbar on the right. We highlight the fraction of galaxies in our sample that reside in each quadrant along with the physical interpretation. These colors correspond to the arrows in the left panel.\textit{(left)} Displaying examples of gradients in the UVJ plane corresponding to the four quadrant in the right panel. The arrows signify the movement from inner to outer radii moving from the base to the head of the arrow. } 
    \label{fig:delta_uvj}
\end{figure*}
\subsection{UVJ Gradients in Star-forming Galaxies}
\label{sec:res_sf}
In this section we focus on the UVJ gradients of star-forming galaxies. They make up roughly 75\% of our sample (39/54) and the physical interpretation of the UVJ plane is well established. Figure~\ref{fig:delta_uvj} displays the differences between $U-V$ and $V-J$ color between the outer galaxy, which we define as $2r_{\rm eff,F356W}<r<3\,r_{\rm eff,F356W}$ and the inner galaxy, defined as $r<r_{\rm eff,F356W}$. Figure~\ref{fig:delta_uvj} visualizes examples from the four quadrants of the $\Delta (U-V) - \Delta(V-J)$ plane and how these gradient would appear in the UVJ color-color plane along with the presumed physical cause of the gradient. We focus on the effects of dust and sSFR, which have orthogonal effects in the UVJ plane. In real galaxies there are likely multiple effects which could cause color gradients, yet these quadrants are a useful tool to understand the dominant physical processes that affect galaxies. 

The largest fraction of star-forming galaxies lie in quadrant three of this plane, corresponding to negative $U-V$ and $V-J$ gradients. The color gradients of these galaxies are consistent with differential dust attenuation, with higher dust content in the center compared to the outskirts. In contrast, only 8\% of galaxies in our sample reside in Q1, where we expect the dust attenuation to be higher in the outskirts. A smaller fraction of galaxies reside in either Q2 or Q4. These gradients are consistent with the following radial variation in stellar populations: high sSFR or younger age in the central region in Q2, or the outskirts for Q4. Galaxies exhibiting a classical old bulge and star-forming disk, like the Milky Way, would likely live somewhere in Q4. These galaxies tend to have relatively high sSFR ($\log\, {\rm sSFR} \sim -9\ \rm yr^{-1}$).

\begin{figure}
    \centering
    \includegraphics[width = 0.99\columnwidth]{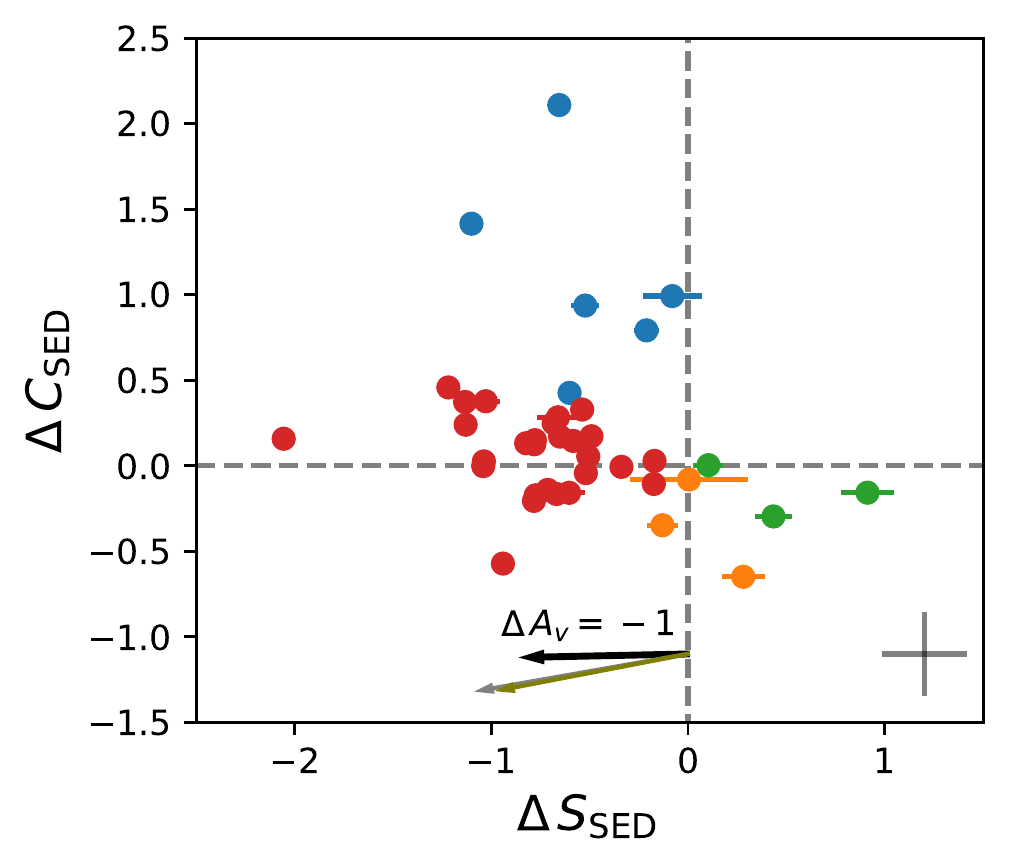}
    \caption{Gradients in $S_{\rm SED}$ and $C_{\rm SED}$ colors, a rotated version of the UVJ color-color space proposed in \citet{Fang2018}. The horizontal axis shows $\Delta, S_{\rm SED}$ which is correlated with dust content and the vertical axis shows $\Delta C_{\rm SED}$, which is correlated with star-formation. Negative values in the horizontal axis imply higher dust attenuation dust in the center of galaxies. The colors of the points represent which quadrant they inhabit in Fig.~\ref{fig:delta_uvj}. The black arrow displays the change in color corresponding to $\Delta A_v = -1$ according to the \citet{Calzetti2000} dust curve. The olive and gray arrow show this instead using the \citet{Reddy2015} and \citep{Salmon2016} (moving from $A_v = 1 \rightarrow 0$) respectively. For the galaxies in Q3 (red points) many of them are consistent with $\Delta C_{\rm SED} = 0$, implying radial decreasing dust attenuation is the sole cause of color gradients.}
    \label{fig:dcol_proj_dust}
\end{figure}

\added{To isolate the effects of dust and sSFR in Figure~\ref{fig:dcol_proj_dust} we show the change in $S_{\rm SED}$ and $C_{\rm SED}$ colors, defined in \citet{Fang2018}. This is a 34.8$^\circ$ rotation of the UVJ axis such that it is parallel with the star-forming sequence. $S_{\rm SED}$ measures the net slope of the SED is correlated with dust content while $C_{\rm SED}$ quantifies the curvature and is correlated with sSSR. Please see \citet{Fang2018} for more details.  For many of the galaxies in Q3, $\Delta\, C_{\rm SED}$ is consistent with 0, implying radially decreasing dust attenuation is the sole cause of the color gradients. There does appear to be a slight bias to positive $\Delta\, C_{\rm SED}$ values, which would imply the additional effect of increased central star-formation.} The implied median change in dust attenuation for galaxies in Q3 is $\Delta A_V = - 0.86$ between the outer and inner regions with a standard deviation of 0.45 $\Delta A_V $ which does not correlate strongly with galaxy properties or redshift.

\subsection{UVJ gradients in Quiescent galaxies}
\label{sec:res_Q}

\begin{figure}
    \centering
    \includegraphics[width = 0.99\columnwidth]{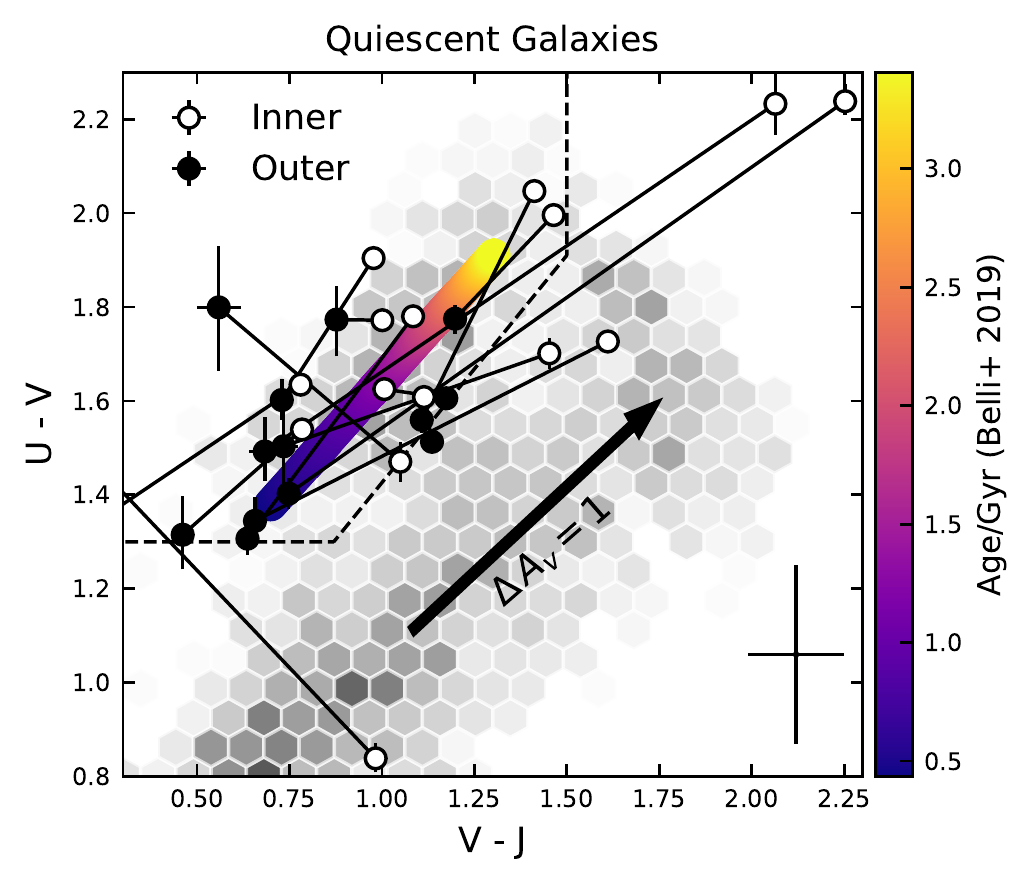}
    \caption{UVJ colors of quiescent galaxies, comparing their inner regions ($r<r_{\rm 356W}$) and outer region ($2\, r_{\rm 356W} < 3\, r_{\rm 356W}$). Galaxies are categorized using their integrated UVJ colors following \citet{Muzzin2013}. The empirical age gradient derived in \citet{Belli2019} is shown along with the \citet{Calzetti2000} dust vector. There is a diversity of UVJ gradients within quiescent galaxies but we note that for roughly one third of our sample (6 of 15) the inner region lies within the star-forming section of the UVJ plane.}
    \label{fig:UVJ_Q}
\end{figure}

Our discussion so far has focused on star-forming galaxies as they make up the majority of our sample and the physical interpretation is relatively straightforward, as described above. The interpretation of quiescent galaxies becomes more complicated as there is known to be an age gradient along this sequence which displays a similar vector to differential dust attenuation \citep{Whitaker2012,Belli2019}. Figure~\ref{fig:UVJ_Q} displays the Calzetti dust vector along with the age sequence measured from \citet{Belli2019}. These two are almost parallel, complicating the interpretation of $\Delta\, (U-V)$ and $\Delta\, (V-J)$ as the galaxy's location in UVJ space becomes important.

Quiescent galaxies in our sample show a range of UVJ gradients. Six of the fifteen galaxies are classified as star-forming in their inner region and within the UVJ quiescent region in their outskirts. \added{Two of these galaxies have very red centers, $U-V >2$ and $V-J > 2$, suggesting high dust attenuation, these galaxies are both quite compact and we do not show any morphological signs of dust} This is similar to the behaviour of star-forming galaxies in Q2 displayed above in Fig.~\ref{fig:delta_uvj}. Another four of the galaxies show relatively mild UVJ gradients within the quiescent region ($\Delta\, (U-V)$ and $\Delta\, (V-J)\ < 0.3$). The final five galaxies show gradients within the quiescent region that are parallel to the age sequence derived in ~\citet{Belli2019} or with radially decreasing dust attenuation.

\section{Discussion and Summary} \label{sec:disc}

In this study we have used early science observations from \emph{JWST} NIRCam to investigate color gradients within galaxies at $z\sim 2$. Specifically, we are interested in the question: what is the underlying cause of color gradients at cosmic noon? Previously, with only rest-frame optical colors at sub-arcsecond resolution, there was no way to definitively answer this question. We use data taken as part of the CEERS survey in the well-studied AEGIS field and construct a sample of galaxies above $\log M_* / M_\odot>10$ at $1.7<z<2.3$ from the 3D-HST catalog. We model the light distribution for our sample of galaxies in three NIRCam bands with \texttt{imcascade}; a Bayesian implementation of the MGE technique which flexibly models galaxy profiles as a mixture of Gaussians. This technique is well-suited for measuring complex galaxy profiles in high SNR data, like those in JWST images, as shown in Figure~\ref{fig:examp_gals}. From the \texttt{imcascade} models we calculate resolved U-V and V-J colors based on the $m_{F115W} - m_{200W}$ and $m_{200W} - m_{F356W}$ colors respectively. 

For star-forming galaxies we observe the large majority of galaxies, roughly 70\%, show UVJ gradients consistent with strong central dust attenuation ($A_V > 1$). Radial dust attenuation gradients have been observed at $z\lesssim 1.5$. \citet{Wang2017} study stacked UVI (Similar to the UVJ plane) gradients in star-forming galaxies out to $z=1.5$ and conclude dust is the main cause of color gradients. \citet{Liu2016} infer dust is the main cause of NUV-B gradients in $0.5<z<1.5$. \citet{Nelson2016} study resolved Balmer decrements and find that the attenuation of the $H_{\alpha}$ line can increase by up to 2 mag in the center of massive galaxies. It has also been suggested that dust gradients continue to play a large role in galaxies at $z>1.5$ using lower wavelength HST data~\citep{Liu2017, Miller2022}. This study provides the first definitive evidence confirming that dust is the cause of negative color gradients out to $z=2.3$. Given that the ratio of FIR and mm sizes continue to be smaller than optical sizes out to $z\sim4$~\citep{Fujimoto2017, Tadakai2020}, we suspect dust gradients will still play a large role in shaping color gradients and observed morphologies in galaxies at higher redshift. Numerical simulations suggest that even at $z>6$ dust continues to play a large role in shaping the rest-frame optical and UV morphology of galaxies \citep{Marshall2022}.

It is worth highlighting this result is qualitatively different to what is observed in spiral galaxies in the local universe. Galaxies like Andromeda or the Milky Way also show negative color gradients, however it is caused by an old, mostly dust-free bulge and a younger star-forming disk \citep{deJong1996}. This would manifest as a UVJ gradient in quadrant four (Orange color in Fig~\ref{fig:delta_uvj}), of which only a small fraction of our sample resides. Our observations are consistent with the scenario of dust-obscured bulge growth at high redshift ~\citep{Tacchella2018, Nelson2019}. It is possible there are also sSFR/age gradients in our sample that are simply being ``outshone'' by the dominant gradient in dust opacity. Resolved SED modelling would be required to further investigate simultaneous gradients of multiple physical properties. Metallicity gradients are also thought to play a large role in local galaxies~\citep[e.g.][]{Tortora2011}. There is not a clear signature of metallicity in the UVJ plane. Again detailed SED modelling or resolved spectroscopy is needed to investigate metallicity gradients.

We find that a substantial fraction of our star-forming galaxies ($23\%$) have UVJ gradients consistent with gradients in stellar population properties, mainly sSFR. These gradients are orthogonal to the dust vector in the UVJ plane. Roughly 2/3 of these galaxies show a central star-formation burst while the other one third show stronger SF in the outskirts. This is combined with a similar of population of quiescent galaxies whose inner regions reside in the star-forming region of UVJ space. While a smaller fraction of the total population of SF galaxies, they likely represent important transitional phases in their formation history. Those with central star-formation could be related to the growth of bulges observed in local spiral galaxies \citep{Nelson2018,Tadaki2020}. The galaxies with outer star-formation could be in the process of inside-out quenching\citep[e.g.][]{Spilker2019, Akhshik2022}. Further studies of these transitional galaxies will help illuminate quenching mechanisms at cosmic noon.

Interpreting UVJ gradients in quiescent galaxies is more complicated. We see some galaxies with color gradients consistent with the known age gradient along the quiescent sequence~\citep{Whitaker2012,Belli2019}, however this is almost parallel to the effect of radially decreasing dust. Other studies have found that UVJ colors alone cannot constrain the age of a stellar population, suggesting instead that observed correlations between UVJ colors and age arise from secondary correlations or scaling relationships with other parameters~\citep{Leja2019}. Resolved spectroscopy or MIR/FIR measurements could help differentiate between the effects of dust and stellar age. The globally quiescent galaxies with ongoing star-formation in their centers may represent an early stage of quenching or experiencing a rejuvination event~\citep[e.g.][]{Akhshik2021}. The presence of quiescent galaxies without strong color gradients provides a clue that there are multiple quenching pathways ~\citep{Woo2019,Suess2021,Akhshik2022}. A larger sample and more detailed modeling is need to fully understand color gradients in early quiescent galaxies.

This study represents the very beginning of how \emph{JWST} will unveil the resolved structure of galaxies at cosmic noon. With rest-frame NIR imaging at the resolution of \emph{JWST}, we will be able to investigate gradients in stellar age, dust an other physical properties at the peak epoch of galaxy growth. \emph{JWST} opens the window to a more complete and detailed view of how galaxies form and evolve in the early universe.

\begin{acknowledgments}
TBM would like to thank Patricia Gruber and the Gruber foundation for their support of work presented here. Cloud-based data processing and file storage for this work is provided by the AWS Cloud Credits for Research program.

Some of the data presented in this paper were obtained from the Mikulski Archive for Space Telescopes (MAST) at the Space Telescope Science Institute. The specific observations analyzed can be accessed via \dataset[10.17909/2hft-5f48]{https://doi.org/10.17909/2hft-5f48}
\end{acknowledgments}

\bibliography{r_UVJ}

\begin{thebibliography}{}
\expandafter\ifx\csname natexlab\endcsname\relax\def\natexlab#1{#1}\fi
\providecommand{\url}[1]{\href{#1}{#1}}
\providecommand{\dodoi}[1]{doi:~\href{http://doi.org/#1}{\nolinkurl{#1}}}
\providecommand{\doeprint}[1]{\href{http://ascl.net/#1}{\nolinkurl{http://ascl.net/#1}}}
\providecommand{\doarXiv}[1]{\href{https://arxiv.org/abs/#1}{\nolinkurl{https://arxiv.org/abs/#1}}}

\bibitem[{{Akhshik} {et~al.}(2021){Akhshik}, {Whitaker}, {Leja}, {Mahler},
  {Sharon}, {Brammer}, {Toft}, {Bezanson}, {Man}, {Nelson}, {Pacifici},
  {Wellons}, \& {Williams}}]{Akhshik2021}
{Akhshik}, M., {Whitaker}, K.~E., {Leja}, J., {et~al.} 2021, \apjl, 907, L8,
  \dodoi{10.3847/2041-8213/abd416}

\bibitem[{{Akhshik} {et~al.}(2022){Akhshik}, {Whitaker}, {Leja}, {Richard},
  {Spilker}, {Song}, {Brammer}, {Bezanson}, {Ebeling}, {Gallazzi}, {Mahler},
  {Mowla}, {Nelson}, {Pacifici}, {Sharon}, {Toft}, {Williams}, {Wright}, \&
  {Zabl}}]{Akhshik2022}
---. 2022, arXiv e-prints, arXiv:2203.04979.
\newblock \doarXiv{2203.04979}

\bibitem[{{Akins} {et~al.}(2022){Akins}, {Narayanan}, {Whitaker}, {Dav{\'e}},
  {Lower}, {Bezanson}, {Feldmann}, \& {Kriek}}]{Akins2022}
{Akins}, H.~B., {Narayanan}, D., {Whitaker}, K.~E., {et~al.} 2022, \apj, 929,
  94, \dodoi{10.3847/1538-4357/ac5d3a}

\bibitem[{{Antwi-Danso} {et~al.}(2022){Antwi-Danso}, {Papovich}, {Leja},
  {Marchesini}, {Marsan}, {Martis}, {Labb{\'e}}, {Muzzin}, {Glazebrook},
  {Straatman}, \& {Tran}}]{AntwiDanso2022}
{Antwi-Danso}, J., {Papovich}, C., {Leja}, J., {et~al.} 2022, arXiv e-prints,
  arXiv:2207.07170.
\newblock \doarXiv{2207.07170}

\bibitem[{{Arnouts} {et~al.}(2013){Arnouts}, {Le Floc'h}, {Chevallard},
  {Johnson}, {Ilbert}, {Treyer}, {Aussel}, {Capak}, {Sanders}, {Scoville},
  {McCracken}, {Milliard}, {Pozzetti}, \& {Salvato}}]{Arnouts2013}
{Arnouts}, S., {Le Floc'h}, E., {Chevallard}, J., {et~al.} 2013, \aap, 558,
  A67, \dodoi{10.1051/0004-6361/201321768}

\bibitem[{{Belli} {et~al.}(2019){Belli}, {Newman}, \& {Ellis}}]{Belli2019}
{Belli}, S., {Newman}, A.~B., \& {Ellis}, R.~S. 2019, \apj, 874, 17,
  \dodoi{10.3847/1538-4357/ab07af}

\bibitem[{{Bezanson} {et~al.}(2016){Bezanson}, {Wake}, {Brammer}, {van Dokkum},
  {Franx}, {Labb{\'e}}, {Leja}, {Momcheva}, {Nelson}, {Quadri}, {Skelton},
  {Weiner}, \& {Whitaker}}]{Bezanson2016}
{Bezanson}, R., {Wake}, D.~A., {Brammer}, G.~B., {et~al.} 2016, \apj, 822, 30,
  \dodoi{10.3847/0004-637X/822/1/30}

\bibitem[{{Boyer} {et~al.}(2022){Boyer}, {Anderson}, {Gennaro}, {Geha},
  {Wingfield McQuinn}, {Tollerud}, {Correnti}, {Brenner Newman}, {Cohen},
  {Kallivayalil}, {Beaton}, {Cole}, {Dolphin}, {Kalirai}, {Sandstrom},
  {Savino}, {Skillman}, {Weisz}, \& {Williams}}]{Boyer2022}
{Boyer}, M.~L., {Anderson}, J., {Gennaro}, M., {et~al.} 2022, arXiv e-prints,
  arXiv:2209.03348.
\newblock \doarXiv{2209.03348}

\bibitem[{Bradley {et~al.}(2020)Bradley, Sip{\H o}cz, Robitaille, Tollerud,
  Vin{\'{\i}}cius, Deil, Barbary, Wilson, Busko, G{\"u}nther, Cara, Conseil,
  Bostroem, Droettboom, Bray, Bratholm, Lim, Barentsen, Craig, Pascual, Perren,
  Greco, Donath, de~Val-Borro, Kerzendorf, Bach, Weaver, D'Eugenio, Souchereau,
  \& Ferreira}]{photutils}
Bradley, L., Sip{\H o}cz, B., Robitaille, T., {et~al.} 2020, astropy/photutils:
  1.0.0, 1.0.0,  Zenodo, \dodoi{10.5281/zenodo.4044744}

\bibitem[{{Brammer}(2019)}]{grizli}
{Brammer}, G. 2019, {Grizli: Grism redshift and line analysis software},
  Astrophysics Source Code Library, record ascl:1905.001.
\newblock \doeprint{1905.001}

\bibitem[{{Brammer} {et~al.}(2008){Brammer}, {van Dokkum}, \&
  {Coppi}}]{Brammer2008}
{Brammer}, G.~B., {van Dokkum}, P.~G., \& {Coppi}, P. 2008, \apj, 686, 1503,
  \dodoi{10.1086/591786}

\bibitem[{{Brammer} {et~al.}(2009){Brammer}, {Whitaker}, {van Dokkum},
  {Marchesini}, {Labb{\'e}}, {Franx}, {Kriek}, {Quadri}, {Illingworth}, {Lee},
  {Muzzin}, \& {Rudnick}}]{Brammer2009}
{Brammer}, G.~B., {Whitaker}, K.~E., {van Dokkum}, P.~G., {et~al.} 2009, \apjl,
  706, L173, \dodoi{10.1088/0004-637X/706/1/L173}

\bibitem[{{Calzetti} {et~al.}(2000){Calzetti}, {Armus}, {Bohlin}, {Kinney},
  {Koornneef}, \& {Storchi-Bergmann}}]{Calzetti2000}
{Calzetti}, D., {Armus}, L., {Bohlin}, R.~C., {et~al.} 2000, \apj, 533, 682,
  \dodoi{10.1086/308692}

\bibitem[{{Chan} {et~al.}(2016){Chan}, {Beifiori}, {Mendel}, {Saglia},
  {Bender}, {Fossati}, {Galametz}, {Wegner}, {Wilman}, {Cappellari}, {Davies},
  {Houghton}, {Prichard}, {Lewis}, {Sharples}, \& {Stott}}]{Chan2016}
{Chan}, J. C.~C., {Beifiori}, A., {Mendel}, J.~T., {et~al.} 2016, \mnras, 458,
  3181, \dodoi{10.1093/mnras/stw502}

\bibitem[{{Davidzon} {et~al.}(2017){Davidzon}, {Ilbert}, {Laigle}, {Coupon},
  {McCracken}, {Delvecchio}, {Masters}, {Capak}, {Hsieh}, {Le F{\`e}vre},
  {Tresse}, {Bethermin}, {Chang}, {Faisst}, {Le Floc'h}, {Steinhardt}, {Toft},
  {Aussel}, {Dubois}, {Hasinger}, {Salvato}, {Sanders}, {Scoville}, \&
  {Silverman}}]{Davidzon2017}
{Davidzon}, I., {Ilbert}, O., {Laigle}, C., {et~al.} 2017, \aap, 605, A70,
  \dodoi{10.1051/0004-6361/201730419}

\bibitem[{{de Jong}(1996)}]{deJong1996}
{de Jong}, R.~S. 1996, \aap, 313, 377.
\newblock \doarXiv{astro-ph/9604010}

\bibitem[{{Fang} {et~al.}(2018){Fang}, {Faber}, {Koo}, {Rodr{\'\i}guez-Puebla},
  {Guo}, {Barro}, {Behroozi}, {Brammer}, {Chen}, {Dekel}, {Ferguson},
  {Gawiser}, {Giavalisco}, {Kartaltepe}, {Kocevski}, {Koekemoer}, {McGrath},
  {McIntosh}, {Newman}, {Pacifici}, {Pandya}, {P{\'e}rez-Gonz{\'a}lez},
  {Primack}, {Salmon}, {Trump}, {Weiner}, {Willner}, {Acquaviva}, {Dahlen},
  {Finkelstein}, {Finlator}, {Fontana}, {Galametz}, {Grogin}, {Gruetzbauch},
  {Johnson}, {Mobasher}, {Papovich}, {Pforr}, {Salvato}, {Santini}, {van der
  Wel}, {Wiklind}, \& {Wuyts}}]{Fang2018}
{Fang}, J.~J., {Faber}, S.~M., {Koo}, D.~C., {et~al.} 2018, \apj, 858, 100,
  \dodoi{10.3847/1538-4357/aabcba}

\bibitem[{{Finkelstein} {et~al.}(2017){Finkelstein}, {Dickinson}, {Ferguson},
  {Grazian}, {Grogin}, {Kartaltepe}, {Kewley}, {Kocevski}, {Koekemoer}, {Lotz},
  {Papovich}, {Pentericci}, {Perez-Gonzalez}, {Pirzkal}, {Ravindranath},
  {Somerville}, {Trump}, \& {Wilkins}}]{Finkelstein2017}
{Finkelstein}, S.~L., {Dickinson}, M., {Ferguson}, H.~C., {et~al.} 2017, {The
  Cosmic Evolution Early Release Science (CEERS) Survey}, JWST Proposal ID
  1345. Cycle 0 Early Release Science

\bibitem[{{Finkelstein} {et~al.}(2022){Finkelstein}, {Bagley}, {Arrabal Haro},
  {Dickinson}, {Ferguson}, {Kartaltepe}, {Papovich}, {Burgarella}, {Kocevski},
  {Huertas-Company}, {Iyer}, {Larson}, {P{\'e}rez-Gonz{\'a}lez}, {Rose},
  {Tacchella}, {Wilkins}, {Chworowsky}, {Medrano}, {Morales}, {Somerville},
  {Yung}, {Fontana}, {Giavalisco}, {Grazian}, {Grogin}, {Kewley}, {Koekemoer},
  {Kirkpatrick}, {Kurczynski}, {Lotz}, {Pentericci}, {Pirzkal}, {Ravindranath},
  {Ryan}, {Trump}, {Yang}, {Almaini}, {Amor{\'\i}n}, {Annunziatella},
  {Backhaus}, {Barro}, {Behroozi}, {Bell}, {Bhatawdekar}, {Bisigello}, {Bromm},
  {Buat}, {Buitrago}, {Calabr{\'o}}, {Casey}, {Castellano}, {Ch{\'a}vez Ortiz},
  {Ciesla}, {Cleri}, {Cohen}, {Cole}, {Cooke}, {Cooper}, {Cooray}, {Costantin},
  {Cox}, {Croton}, {Daddi}, {Dav{\'e}}, {de la Vega}, {Dekel}, {Elbaz},
  {Estrada-Carpenter}, {Faber}, {Fern{\'a}ndez}, {Finkelstein}, {Freundlich},
  {Fujimoto}, {Garc{\'\i}a-Argum{\'a}nez}, {Gardner}, {Gawiser},
  {G{\'o}mez-Guijarro}, {Guo}, {Hamilton}, {Hathi}, {Holwerda}, {Hirschmann},
  {Hutchison}, {Jha}, {Jogee}, {Juneau}, {Jung}, {Kassin}, {Le Bail}, {Leung},
  {Lucas}, {Magnelli}, {Mantha}, {Matharu}, {McGrath}, {McIntosh}, {Merlin},
  {Mobasher}, {Newman}, {Nicholls}, {Pandya}, {Rafelski}, {Ronayne}, {Santini},
  {Seill{\'e}}, {Shah}, {Shen}, {Simons}, {Snyder}, {Stanway}, {Straughn},
  {Teplitz}, {Vanderhoof}, {Vega-Ferrero}, {Wang}, {Weiner}, {Willmer},
  {Wuyts}, \& {Zavala}}]{Finkelstein2022}
{Finkelstein}, S.~L., {Bagley}, M.~B., {Arrabal Haro}, P., {et~al.} 2022, arXiv
  e-prints, arXiv:2207.12474.
\newblock \doarXiv{2207.12474}

\bibitem[{{Franx} \& {Illingworth}(1990)}]{Franx1990}
{Franx}, M., \& {Illingworth}, G. 1990, \apjl, 359, L41, \dodoi{10.1086/185791}

\bibitem[{{Fujimoto} {et~al.}(2017){Fujimoto}, {Ouchi}, {Shibuya}, \&
  {Nagai}}]{Fujimoto2017}
{Fujimoto}, S., {Ouchi}, M., {Shibuya}, T., \& {Nagai}, H. 2017, \apj, 850, 83,
  \dodoi{10.3847/1538-4357/aa93e6}

\bibitem[{{Ilbert} {et~al.}(2013){Ilbert}, {McCracken}, {Le F{\`e}vre},
  {Capak}, {Dunlop}, {Karim}, {Renzini}, {Caputi}, {Boissier}, {Arnouts},
  {Aussel}, {Comparat}, {Guo}, {Hudelot}, {Kartaltepe}, {Kneib}, {Krogager},
  {Le Floc'h}, {Lilly}, {Mellier}, {Milvang-Jensen}, {Moutard}, {Onodera},
  {Richard}, {Salvato}, {Sanders}, {Scoville}, {Silverman}, {Taniguchi},
  {Tasca}, {Thomas}, {Toft}, {Tresse}, {Vergani}, {Wolk}, \&
  {Zirm}}]{Ilbert2013}
{Ilbert}, O., {McCracken}, H.~J., {Le F{\`e}vre}, O., {et~al.} 2013, \aap, 556,
  A55, \dodoi{10.1051/0004-6361/201321100}

\bibitem[{{Johnson} {et~al.}(2021){Johnson}, {Leja}, {Conroy}, \&
  {Speagle}}]{Johnson2021}
{Johnson}, B.~D., {Leja}, J., {Conroy}, C., \& {Speagle}, J.~S. 2021, \apjs,
  254, 22, \dodoi{10.3847/1538-4365/abef67}

\bibitem[{{Kormendy} \& {Djorgovski}(1989)}]{Kormendy1989}
{Kormendy}, J., \& {Djorgovski}, S. 1989, \araa, 27, 235,
  \dodoi{10.1146/annurev.aa.27.090189.001315}

\bibitem[{{Kriek} \& {Conroy}(2013)}]{Kriek2013}
{Kriek}, M., \& {Conroy}, C. 2013, \apjl, 775, L16,
  \dodoi{10.1088/2041-8205/775/1/L16}

\bibitem[{{La Barbera} \& {de Carvalho}(2009)}]{LaBarbera2009}
{La Barbera}, F., \& {de Carvalho}, R.~R. 2009, \apjl, 699, L76,
  \dodoi{10.1088/0004-637X/699/2/L76}

\bibitem[{{Labb{\'e}} {et~al.}(2005){Labb{\'e}}, {Huang}, {Franx}, {Rudnick},
  {Barmby}, {Daddi}, {van Dokkum}, {Fazio}, {F{\"o}rster Schreiber},
  {Moorwood}, {Rix}, {R{\"o}ttgering}, {Trujillo}, \& {van der
  Werf}}]{Labbe2005}
{Labb{\'e}}, I., {Huang}, J., {Franx}, M., {et~al.} 2005, \apjl, 624, L81,
  \dodoi{10.1086/430700}

\bibitem[{{Leja} {et~al.}(2017){Leja}, {Johnson}, {Conroy}, {van Dokkum}, \&
  {Byler}}]{Leja2017}
{Leja}, J., {Johnson}, B.~D., {Conroy}, C., {van Dokkum}, P.~G., \& {Byler}, N.
  2017, \apj, 837, 170, \dodoi{10.3847/1538-4357/aa5ffe}

\bibitem[{{Leja} {et~al.}(2019){Leja}, {Tacchella}, \& {Conroy}}]{Leja2019}
{Leja}, J., {Tacchella}, S., \& {Conroy}, C. 2019, \apjl, 880, L9,
  \dodoi{10.3847/2041-8213/ab2f8c}

\bibitem[{{Liu} {et~al.}(2016){Liu}, {Jiang}, {Guo}, {Koo}, {Faber}, {Zheng},
  {Yesuf}, {Barro}, {Li}, {Li}, {Wang}, {Mao}, \& {Fang}}]{Liu2016}
{Liu}, F.~S., {Jiang}, D., {Guo}, Y., {et~al.} 2016, \apjl, 822, L25,
  \dodoi{10.3847/2041-8205/822/2/L25}

\bibitem[{{Liu} {et~al.}(2017){Liu}, {Jiang}, {Faber}, {Koo}, {Yesuf},
  {Tacchella}, {Mao}, {Wang}, {Guo}, {Fang}, {Barro}, {Zheng}, {Jia}, {Tong},
  {Liu}, \& {Meng}}]{Liu2017}
{Liu}, F.~S., {Jiang}, D., {Faber}, S.~M., {et~al.} 2017, \apjl, 844, L2,
  \dodoi{10.3847/2041-8213/aa7cf5}

\bibitem[{{Marshall} {et~al.}(2022){Marshall}, {Wilkins}, {Di Matteo}, {Roper},
  {Vijayan}, {Ni}, {Feng}, \& {Croft}}]{Marshall2022}
{Marshall}, M.~A., {Wilkins}, S., {Di Matteo}, T., {et~al.} 2022, \mnras, 511,
  5475, \dodoi{10.1093/mnras/stac380}

\bibitem[{{Miller} \& {van Dokkum}(2021)}]{Miller2021}
{Miller}, T.~B., \& {van Dokkum}, P. 2021, \apj, 923, 124,
  \dodoi{10.3847/1538-4357/ac2b30}

\bibitem[{{Miller} {et~al.}(2022){Miller}, {van Dokkum}, \&
  {Mowla}}]{Miller2022}
{Miller}, T.~B., {van Dokkum}, P., \& {Mowla}, L. 2022, arXiv e-prints,
  arXiv:2207.05895.
\newblock \doarXiv{2207.05895}

\bibitem[{{Momcheva} {et~al.}(2016){Momcheva}, {Brammer}, {van Dokkum},
  {Skelton}, {Whitaker}, {Nelson}, {Fumagalli}, {Maseda}, {Leja}, {Franx},
  {Rix}, {Bezanson}, {Da Cunha}, {Dickey}, {F{\"o}rster Schreiber},
  {Illingworth}, {Kriek}, {Labb{\'e}}, {Ulf Lange}, {Lundgren}, {Magee},
  {Marchesini}, {Oesch}, {Pacifici}, {Patel}, {Price}, {Tal}, {Wake}, {van der
  Wel}, \& {Wuyts}}]{Momcheva2016}
{Momcheva}, I.~G., {Brammer}, G.~B., {van Dokkum}, P.~G., {et~al.} 2016, \apjs,
  225, 27, \dodoi{10.3847/0067-0049/225/2/27}

\bibitem[{{Mosleh} {et~al.}(2020){Mosleh}, {Hosseinnejad},
  {Hosseini-ShahiSavandi}, \& {Tacchella}}]{Mosleh2020}
{Mosleh}, M., {Hosseinnejad}, S., {Hosseini-ShahiSavandi}, S.~Z., \&
  {Tacchella}, S. 2020, \apj, 905, 170, \dodoi{10.3847/1538-4357/abc7cc}

\bibitem[{{Muzzin} {et~al.}(2013){Muzzin}, {Marchesini}, {Stefanon}, {Franx},
  {McCracken}, {Milvang-Jensen}, {Dunlop}, {Fynbo}, {Brammer}, {Labb{\'e}}, \&
  {van Dokkum}}]{Muzzin2013}
{Muzzin}, A., {Marchesini}, D., {Stefanon}, M., {et~al.} 2013, \apj, 777, 18,
  \dodoi{10.1088/0004-637X/777/1/18}

\bibitem[{{Nelson} {et~al.}(2018){Nelson}, {Pillepich}, {Springel},
  {Weinberger}, {Hernquist}, {Pakmor}, {Genel}, {Torrey}, {Vogelsberger},
  {Kauffmann}, {Marinacci}, \& {Naiman}}]{Nelson2018}
{Nelson}, D., {Pillepich}, A., {Springel}, V., {et~al.} 2018, \mnras, 475, 624,
  \dodoi{10.1093/mnras/stx3040}

\bibitem[{{Nelson} {et~al.}(2016){Nelson}, {van Dokkum}, {F{\"o}rster
  Schreiber}, {Franx}, {Brammer}, {Momcheva}, {Wuyts}, {Whitaker}, {Skelton},
  {Fumagalli}, {Hayward}, {Kriek}, {Labb{\'e}}, {Leja}, {Rix}, {Tacconi}, {van
  der Wel}, {van den Bosch}, {Oesch}, {Dickey}, \& {Ulf Lange}}]{Nelson2016}
{Nelson}, E.~J., {van Dokkum}, P.~G., {F{\"o}rster Schreiber}, N.~M., {et~al.}
  2016, \apj, 828, 27, \dodoi{10.3847/0004-637X/828/1/27}

\bibitem[{Nelson {et~al.}(2019)Nelson, Tadaki, Tacconi, Lutz, Schreiber,
  Cibinel, Wuyts, Lang, Leja, Montes, {et~al.}}]{Nelson2019}
Nelson, E.~J., Tadaki, K.-i., Tacconi, L.~J., {et~al.} 2019, The Astrophysical
  Journal, 870, 130

\bibitem[{{Nelson} {et~al.}(2022){Nelson}, {Suess}, {Bezanson}, {Price}, {van
  Dokkum}, {Leja}, {Whitaker}, {Labb{\'e}}, {Barrufet}, {Brammer},
  {Eisenstein}, {Heintz}, {Johnson}, {Mathews}, {Miller}, {Oesch}, {Sandles},
  {Setton}, {Speagle}, {Tacchella}, {Tadaki}, \& {Weaver}}]{Nelson2022}
{Nelson}, E.~J., {Suess}, K.~A., {Bezanson}, R., {et~al.} 2022, arXiv e-prints,
  arXiv:2208.01630.
\newblock \doarXiv{2208.01630}

\bibitem[{{Patel} {et~al.}(2012){Patel}, {Holden}, {Kelson}, {Franx}, {van der
  Wel}, \& {Illingworth}}]{Patel2012}
{Patel}, S.~G., {Holden}, B.~P., {Kelson}, D.~D., {et~al.} 2012, \apjl, 748,
  L27, \dodoi{10.1088/2041-8205/748/2/L27}

\bibitem[{{Perrin} {et~al.}(2014){Perrin}, {Sivaramakrishnan}, {Lajoie},
  {Elliott}, {Pueyo}, {Ravindranath}, \& {Albert}}]{webbpsf}
{Perrin}, M.~D., {Sivaramakrishnan}, A., {Lajoie}, C.-P., {et~al.} 2014, in
  Society of Photo-Optical Instrumentation Engineers (SPIE) Conference Series,
  Vol. 9143, Space Telescopes and Instrumentation 2014: Optical, Infrared, and
  Millimeter Wave, ed. J.~{Oschmann}, Jacobus~M., M.~{Clampin}, G.~G. {Fazio},
  \& H.~A. {MacEwen}, 91433X, \dodoi{10.1117/12.2056689}

\bibitem[{{Reddy} {et~al.}(2015){Reddy}, {Kriek}, {Shapley}, {Freeman},
  {Siana}, {Coil}, {Mobasher}, {Price}, {Sanders}, \& {Shivaei}}]{Reddy2015}
{Reddy}, N.~A., {Kriek}, M., {Shapley}, A.~E., {et~al.} 2015, \apj, 806, 259,
  \dodoi{10.1088/0004-637X/806/2/259}

\bibitem[{{Salmon} {et~al.}(2016){Salmon}, {Papovich}, {Long}, {Willner},
  {Finkelstein}, {Ferguson}, {Dickinson}, {Duncan}, {Faber}, {Hathi},
  {Koekemoer}, {Kurczynski}, {Newman}, {Pacifici}, {P{\'e}rez-Gonz{\'a}lez}, \&
  {Pforr}}]{Salmon2016}
{Salmon}, B., {Papovich}, C., {Long}, J., {et~al.} 2016, \apj, 827, 20,
  \dodoi{10.3847/0004-637X/827/1/20}

\bibitem[{{Skelton} {et~al.}(2014){Skelton}, {Whitaker}, {Momcheva}, {Brammer},
  {van Dokkum}, {Labb{\'e}}, {Franx}, {van der Wel}, {Bezanson}, {Da Cunha},
  {Fumagalli}, {F{\"o}rster Schreiber}, {Kriek}, {Leja}, {Lundgren}, {Magee},
  {Marchesini}, {Maseda}, {Nelson}, {Oesch}, {Pacifici}, {Patel}, {Price},
  {Rix}, {Tal}, {Wake}, \& {Wuyts}}]{Skelton2014}
{Skelton}, R.~E., {Whitaker}, K.~E., {Momcheva}, I.~G., {et~al.} 2014, \apjs,
  214, 24, \dodoi{10.1088/0067-0049/214/2/24}

\bibitem[{{Speagle}(2020)}]{Speagle2020}
{Speagle}, J.~S. 2020, \mnras, 493, 3132, \dodoi{10.1093/mnras/staa278}

\bibitem[{{Spilker} {et~al.}(2019){Spilker}, {Bezanson}, {Weiner}, {Whitaker},
  \& {Williams}}]{Spilker2019}
{Spilker}, J.~S., {Bezanson}, R., {Weiner}, B.~J., {Whitaker}, K.~E., \&
  {Williams}, C.~C. 2019, \apj, 883, 81, \dodoi{10.3847/1538-4357/ab3804}

\bibitem[{{Suess} {et~al.}(2019){Suess}, {Kriek}, {Price}, \&
  {Barro}}]{Suess2019a}
{Suess}, K.~A., {Kriek}, M., {Price}, S.~H., \& {Barro}, G. 2019, \apj, 877,
  103, \dodoi{10.3847/1538-4357/ab1bda}

\bibitem[{{Suess} {et~al.}(2021){Suess}, {Kriek}, {Price}, \&
  {Barro}}]{Suess2021}
---. 2021, \apj, 915, 87, \dodoi{10.3847/1538-4357/abf1e4}

\bibitem[{{Szomoru} {et~al.}(2013){Szomoru}, {Franx}, {van Dokkum}, {Trenti},
  {Illingworth}, {Labb{\'e}}, \& {Oesch}}]{Szomoru2013}
{Szomoru}, D., {Franx}, M., {van Dokkum}, P.~G., {et~al.} 2013, \apj, 763, 73,
  \dodoi{10.1088/0004-637X/763/2/73}

\bibitem[{Tacchella {et~al.}(2018)Tacchella, Carollo, Schreiber, Renzini,
  Dekel, Genzel, Lang, Lilly, Mancini, Onodera, {et~al.}}]{Tacchella2018}
Tacchella, S., Carollo, C.~M., Schreiber, N.~F., {et~al.} 2018, The
  Astrophysical Journal, 859, 56

\bibitem[{{Tadaki} {et~al.}(2020{\natexlab{a}}){Tadaki}, {Belli}, {Burkert},
  {Dekel}, {F{\"o}rster Schreiber}, {Genzel}, {Hayashi}, {Herrera-Camus},
  {Kodama}, {Kohno}, {Koyama}, {Lee}, {Lutz}, {Mowla}, {Nelson}, {Renzini},
  {Suzuki}, {Tacconi}, {{\"U}bler}, {Wisnioski}, \& {Wuyts}}]{Tadakai2020}
{Tadaki}, K.-i., {Belli}, S., {Burkert}, A., {et~al.} 2020{\natexlab{a}}, \apj,
  901, 74, \dodoi{10.3847/1538-4357/abaf4a}

\bibitem[{{Tadaki} {et~al.}(2020{\natexlab{b}}){Tadaki}, {Belli}, {Burkert},
  {Dekel}, {F{\"o}rster Schreiber}, {Genzel}, {Hayashi}, {Herrera-Camus},
  {Kodama}, {Kohno}, {Koyama}, {Lee}, {Lutz}, {Mowla}, {Nelson}, {Renzini},
  {Suzuki}, {Tacconi}, {{\"U}bler}, {Wisnioski}, \& {Wuyts}}]{Tadaki2020}
---. 2020{\natexlab{b}}, \apj, 901, 74, \dodoi{10.3847/1538-4357/abaf4a}

\bibitem[{{Tortora} {et~al.}(2011){Tortora}, {Napolitano}, {Romanowsky},
  {Jetzer}, {Cardone}, \& {Capaccioli}}]{Tortora2011}
{Tortora}, C., {Napolitano}, N.~R., {Romanowsky}, A.~J., {et~al.} 2011, \mnras,
  418, 1557, \dodoi{10.1111/j.1365-2966.2011.19438.x}

\bibitem[{{van der Wel} {et~al.}(2012){van der Wel}, {Bell}, {H{\"a}ussler},
  {McGrath}, {Chang}, {Guo}, {McIntosh}, {Rix}, {Barden}, {Cheung}, {Faber},
  {Ferguson}, {Galametz}, {Grogin}, {Hartley}, {Kartaltepe}, {Kocevski},
  {Koekemoer}, {Lotz}, {Mozena}, {Peth}, \& {Peng}}]{vanderWel2012}
{van der Wel}, A., {Bell}, E.~F., {H{\"a}ussler}, B., {et~al.} 2012, \apjs,
  203, 24, \dodoi{10.1088/0067-0049/203/2/24}

\bibitem[{{Wang} {et~al.}(2017){Wang}, {Faber}, {Liu}, {Guo}, {Pacifici},
  {Koo}, {Kassin}, {Mao}, {Fang}, {Chen}, {Koekemoer}, {Kocevski}, \&
  {Ashby}}]{Wang2017}
{Wang}, W., {Faber}, S.~M., {Liu}, F.~S., {et~al.} 2017, \mnras, 469, 4063,
  \dodoi{10.1093/mnras/stx1148}

\bibitem[{{Whitaker} {et~al.}(2012){Whitaker}, {Kriek}, {van Dokkum},
  {Bezanson}, {Brammer}, {Franx}, \& {Labb{\'e}}}]{Whitaker2012}
{Whitaker}, K.~E., {Kriek}, M., {van Dokkum}, P.~G., {et~al.} 2012, \apj, 745,
  179, \dodoi{10.1088/0004-637X/745/2/179}

\bibitem[{{Whitaker} {et~al.}(2011){Whitaker}, {Labb{\'e}}, {van Dokkum},
  {Brammer}, {Kriek}, {Marchesini}, {Quadri}, {Franx}, {Muzzin}, {Williams},
  {Bezanson}, {Illingworth}, {Lee}, {Lundgren}, {Nelson}, {Rudnick}, {Tal}, \&
  {Wake}}]{Whitaker2011}
{Whitaker}, K.~E., {Labb{\'e}}, I., {van Dokkum}, P.~G., {et~al.} 2011, \apj,
  735, 86, \dodoi{10.1088/0004-637X/735/2/86}

\bibitem[{{Whitaker} {et~al.}(2013){Whitaker}, {van Dokkum}, {Brammer},
  {Momcheva}, {Skelton}, {Franx}, {Kriek}, {Labb{\'e}}, {Fumagalli},
  {Lundgren}, {Nelson}, {Patel}, \& {Rix}}]{Whitaker2013}
{Whitaker}, K.~E., {van Dokkum}, P.~G., {Brammer}, G., {et~al.} 2013, \apjl,
  770, L39, \dodoi{10.1088/2041-8205/770/2/L39}

\bibitem[{{Williams} {et~al.}(2009){Williams}, {Quadri}, {Franx}, {van Dokkum},
  \& {Labb{\'e}}}]{Williams2009}
{Williams}, R.~J., {Quadri}, R.~F., {Franx}, M., {van Dokkum}, P., \&
  {Labb{\'e}}, I. 2009, \apj, 691, 1879, \dodoi{10.1088/0004-637X/691/2/1879}

\bibitem[{{Woo} \& {Ellison}(2019)}]{Woo2019}
{Woo}, J., \& {Ellison}, S.~L. 2019, \mnras, 487, 1927,
  \dodoi{10.1093/mnras/stz1377}

\bibitem[{{Wu} {et~al.}(2005){Wu}, {Shao}, {Mo}, {Xia}, \& {Deng}}]{Wu2005}
{Wu}, H., {Shao}, Z., {Mo}, H.~J., {Xia}, X., \& {Deng}, Z. 2005, \apj, 622,
  244, \dodoi{10.1086/427821}

\bibitem[{{Wuyts} {et~al.}(2007){Wuyts}, {Labb{\'e}}, {Franx}, {Rudnick}, {van
  Dokkum}, {Fazio}, {F{\"o}rster Schreiber}, {Huang}, {Moorwood}, {Rix},
  {R{\"o}ttgering}, \& {van der Werf}}]{Wuyts2007}
{Wuyts}, S., {Labb{\'e}}, I., {Franx}, M., {et~al.} 2007, \apj, 655, 51,
  \dodoi{10.1086/509708}

\bibitem[{{Wuyts} {et~al.}(2012){Wuyts}, {F{\"o}rster Schreiber}, {Genzel},
  {Guo}, {Barro}, {Bell}, {Dekel}, {Faber}, {Ferguson}, {Giavalisco}, {Grogin},
  {Hathi}, {Huang}, {Kocevski}, {Koekemoer}, {Koo}, {Lotz}, {Lutz}, {McGrath},
  {Newman}, {Rosario}, {Saintonge}, {Tacconi}, {Weiner}, \& {van der
  Wel}}]{Wuyts2012}
{Wuyts}, S., {F{\"o}rster Schreiber}, N.~M., {Genzel}, R., {et~al.} 2012, \apj,
  753, 114, \dodoi{10.1088/0004-637X/753/2/114}

\bibitem[{{Zuckerman} {et~al.}(2021){Zuckerman}, {Belli}, {Leja}, \&
  {Tacchella}}]{Zuckerman2021}
{Zuckerman}, L.~D., {Belli}, S., {Leja}, J., \& {Tacchella}, S. 2021, \apjl,
  922, L32, \dodoi{10.3847/2041-8213/ac3831}

\end{thebibliography}
\bibliographystyle{aasjournal}

\end{document}